# The decay $b \to s\gamma$ in SUSY extensions of the standard model

Pran Nath[†‡§] and R. Arnowitt[∥]

† Theoretical Physics Division, CERN, CH-1211 Geneva 23, Switzerland
∥ Center for Theoretical Physics, Department of Physics, Texas A & M University, College Station, TX 77843, USA

**Abstract**

A brief review is given of the decay $b \to s\gamma$ in SUSY extensions of the Standard Model. It is found that the recent CLEO results put strong constraints on the parameter space of minimal N=1 Supergravity unified theory. Dark Matter analyses are also strongly constrained for $\mu > 0$.

## 1. Introduction

Recently, the experimental situation on the measurement of $b \to s\gamma$ branching ratio has improved dramatically. Last year the CLEO Collaboration [1] found an upper bound of $B(b \to s\gamma) < 5.4 \times 10^{-4}$ at 95% CL. This result is now superseded by the first actual measurement of this process reported at this conference. Thus CLEO gives [2]

$$B(b \to s\gamma) = (2.32 \pm 0.51 \pm 0.29 \pm 0.32) \times 10^{-4} \quad (1)$$

where the first error is statistical, the second error is systematic arising from uncertainty in yield, and the third error is also systematic arising from uncertainty in efficiency. In this paper we discuss the implications of these results for supersymmetric extensions of the Standard Model which depend very much on the value of the branching ratio predicted by the SM. Thus we begin by reviewing briefly the current status of the SM prediction for the $b \to s\gamma$ decay.

To leading QCD order $B(b \to s\gamma)$ is given by [3]

$$\frac{B(b \to s\gamma)}{B(b \to ce\bar{\nu})} = \frac{6\alpha_{em}}{\pi\rho\lambda} \frac{|V_{ts}^\star V_{tb}|^2}{|V_{cb}|^2} |\bar{c}_7(m_b)|^2 \quad (2a)$$

where $\rho$ is a phase-space factor, $\lambda$ is a QCD correction factor, $V_{ts}$ etc. are KM matrix elements and $\bar{c}_7(m_b)$ is the effective Wilson co-efficient of the photonic magnetic penguin at scale the $m_b$, i.e.,

$$\bar{c}_7(m_b) = \eta^{\frac{16}{23}} c_7(M_W) + \frac{8}{3}\left(\eta^{\frac{14}{23}} - \eta^{\frac{16}{23}}\right) c_8(M_W) + c_2 \quad (2b)$$

‡ Permanent address: Department of Physics, Northeastern University, Boston, MA 02115, USA
§ Speaker at the Conference

Where $\eta = \alpha_s(M_W)/\alpha_s(m_b)$, $c_7(c_8)$ are the Wilson co-efficients for the photonic (gluonic) magnetic penguins at scale $M_W$ and $c_2$ is an operator mixing co-efficient. In the SM, $c_7(c_8)$ receive contributions from $W$-exchange. The evaluation of $c_2$ depends on the computation to $\mathcal{O}(g^2)$ of an $8 \times 8$ anomalous dimension matrix. The previous $\mathcal{O}(1)\%$ ambiguities in this computation have now been resolved as reported by Ciuchini here [4]. The analysis of $B(b \to s\gamma)$ in the SM using equation (2) suffers from many uncertainties. These include experimental uncertainties in the quark masses, in $\alpha_s$, and in the KM matrix elements. However, the largest uncertainty arises due to the possible next-to-leading order QCD corrections. These could be in the vicinity of $\mathcal{O}(30)\%$ or more [5]. Recently Ciuchini et al. have obtained an upgraded theoretical evaluation for $B(b \to s\gamma)$ in SM using all the known (but incomplete) next to leading order (NLO) corrections [4]. They give a value of

$$B(b \to s\gamma) = (1.9 \pm 0.2 \pm 0.5) \times 10^{-4} \quad (4)$$

However, equation (3) is a mean of two significantly different evaluations; one which uses the t' Hooft-Veltman regularization and the second one which uses the naive dimensional reduction regularization. In view of this many workers prefer to use only the leading order (LO) prediction of SM, pending the full NLO evaluation in SM. For example the CLEO Collaboration uses a mean LO SM value of

$$B(b \to s\gamma) = (2.75 \pm 0.8) \times 10^{-4} \quad (5)$$

for comparison of their experimental results with theory. In our analysis we shall choose the range given by equations (3) and (4). The reason for enumerating the uncertainties in the evaluation of $B(b \to s\gamma)$ in the SM is that many of these uncertainties are generic and similar uncertainties appear when one computes the branching ratio in models based on extensions of SM.

There are several ways in which one can carry out a SUSY extension of the Standard Model. These include the minimal extension, and the non-minimal extensions where either there is extra matter, or the gauge group is larger (such as L-R symmetric models) and variations there of [6]. Here we shall discuss only the minimal extension. The minimal SUSY extension (MSSM) consists of adjoining SUSY multiplets to the $SU(3)_c \otimes SU(2)_L \otimes U(1)_Y$ quark-lepton multiplets of the SM and introducing a pair of Higgs doublets and their SUSY partners. Thus in addition to quarks and leptons the additional states consist of 32 SUSY particles (these are 12 squarks, 9 sleptons, 2 charginos, 4 neutralinos, 1 gluino and 4 Higgs). In the MSSM there are additional contributions to $B(b \to s\gamma)$ arising from the exchange of the charged Higgs, the charginos, the neutralinos, the gluino, and the squarks [7]. In all twenty new (supersymmetric) states enter in the analysis. The physics of $b \to s\gamma$ decay is controlled by the mechanism of supersymmetry breaking. This can be understood from the fact that one has a cancellation of $c_7$ and $c_8$ in the exact SUSY limit [8]. Thus the parameters that characterize SUSY breaking are central to the computation of $c_7$ and $c_8$. Unfortunately the MSSM, does not accommodate a phenomenologically viable way of breaking supersymmetry spontaneously. To generate a viable phenomenology one must add soft SUSY breaking terms by hand to the MSSM. However, the number of allowed possibilities is enormous. One can add up to 137 different soft SUSY breaking terms to the theory. A sharp reduction in the number of soft SUSY breaking parameters occurs within the framework of N=1 supergravity grand unification [9]. Coupled with radiative breaking of the electro-weak symmetry the parameter space of the theory becomes 4 dimensional. The conventional choice of the residual parameters is [10] $m_0$, $m_{1/2}$, $A_0$ and $\tan\beta$ where $m_0$ is the universal scalar mass, $m_{1/2}$ is the universal gaugino mass and $A_0$ is the trilinear coupling in the potential that breaks supersymmetry softly. The analysis of Ref. [11] chooses a different residual set of parameters than the ones above. In that analysis $A_0$ is replaced by $B_0$ where $B_0$ the co-efficeint of the Higgs mixing term in the soft SUSY breaking potential.

We give now a brief description of the $b \to s\gamma$ branching ratio in supergravity grand unification. Many analyses of this decay have appeared recently [12-15, 11]. First the contributions of neutralino and gluino exchange are found to be typically small and we neglect these in our analysis. Charged Higgs make contributions which are always constructive relative to the $W$-exchange [12]. However, the chargino exchange contributions are very model dependent and can be either constructive or destructive [13-15]. An interesting phenomenon that surfaces is that $B(b \to s\gamma)$ can become very small even away from the exact SUSY limit due to cancellations among the $W$, charged Higgs and chargino exchange [14-15]. In general the $b \to s\gamma$ experiment constrains the parameter space of N=1 minimal supergravity [14-15]. An important effect relates to the sensitivity of the $b \to s\gamma$ rate in the region when one is close to the Landau pole [14-16]. In this domain small variations in the input parameters such as $m_t$, $\alpha_G$ and $\tan\beta$ can lead to large variations in the output quantites [14-16].

Another interesting phenomenon relates to the effect of the $b \to s\gamma$ experiment on dark matter analyses. The effect of the experimental constraints of CLEO 93 results on $b \to s\gamma$ on analyses of dark matter were investigated in references [15-17]. It was found that the CLEO 93 results put very strong constraints on dark matter for $\mu > 0$. Here $\mathcal{O}(2/3)$ of the parameter space which satisfies dark matter constraints implied by the COBE constraint [18] is eliminated. For $\mu < 0$, the constraints were less stringent in that only $\mathcal{O}(1/5)$ of the parameter space was eliminated. Similar conclusions hold for the CLEO 94 results of equation (1). However, analysis of reference (16) shows that the CLEO 93 bounds do not constrain the minimal $SU(5)$ model very much. A similar results holds for the CLEO 94 result of equation (1).

One convenient way to quantify SUSY effects is via the parameter defined by

$$r_{SUSY} = \frac{B(b \to s\gamma)_{SUSY}}{B(b \to s\gamma)_{SM}} \quad (6)$$

To leading QCD order and ignoring SUSY threshold effects equation (2) also holds for the minimal N=1 supergravity extension with the only difference that $c_7(c_8)$ in equation (2) are modified to include the charged Higgs and superparticle exchanges. In this approximation $r_{SUSY}$ is given by the ratio of $\bar{c}_7(m_b)$ for the SUSY and the SM cases and is thus relatively free of the ambiguities of the outside factors in equation (2a). Setting $B(b \to s\gamma)_{SUSY}$ to the experimental value of equation (1), and using the range of SM values given by equations (3) and (4), we find the following range for

$r_{SUSY}$:

$$r_{SUSY} = (0.46 - 2.2) \quad (7)$$

which has an average value of $r_{SUSY} = 1.33$. An interesting phenomenon is related to the implication of equation (6) for the SUSY spectrum. Figure 1 exhibits the maximum and the minimum values of the low lying SUSY particles (the light Higgs, the light chargino and the light stop) as a function of $r_{max}$ where $r_{SUSY}$ is allowed to vary in the interval $(0.46-r_{max})$ and $r_{max}$ lies in the range given by equation (6). One finds that SUSY mass bands exhibit a significant narrowing as $r_{SUSY}$ falls below 1. This phenomenon arises due to the constraint that one needs a light SUSY spectrum to cancel the effect of the $W$ and charged Higgs exchange and move $r_{SUSY}$ below the canonical SM value of 1.

Figure 1. Mass bands for the light Higgs (dash-dot), chargino (dashed) and the light stop (solid) as a function of $r_{max}$ when $\mu > 0$, $m_t = 168$ GeV and all other parameters are integrated out.

In conclusion, the CLEO results on $b \to s\gamma$ put severe constraints on the parameter space of minimal supergravity and also significantly affect SUSY dark matter analyses. Specifically it is found that the neutralino relic density analysis for the case $\mu > 0$ is significantly affected. Also the maximum event rates for the detection of neutralinos in dark matter detectors are reduced for the $\mu > 0$ case. However, discovery of supergravity via $b \to s\gamma$ decay would require the full analysis of NLO corrections in supergravity theory including threshold corrections in the evolution of Wilson co-efficients due to different SUSY [19] thresholds, as well as significant further improvement in experiment.